# Memory Efficient Multithreaded Incremental Segmented Sieve Algorithm

Evan Ning[1], David R. Kaeli[2]

Milton Academy[1]
evan_ning24@milton.edu

Department of Electrical and Computer Engineering[2]
Northeastern University
kaeli@ece.neu.edu

## 1. Introduction

Prime numbers are fundamental in number theory and play a significant role in various areas, from pure mathematics to practical applications. For example, public key encryption techniques, especially in cryptography (e.g., the Rivest Shamir Adelman encryption scheme (RSA) [1]), rely heavily on the novel properties of prime numbers. The degree of difficulty in factorizing very large numbers which are products of two prime numbers ensures the security of data encrypted using these methods.

Even though prime numbers are fundamental in number theory, determining the primality of large numbers or generating long lists of prime numbers can be computationally challenging as primes are pseudo-randomly distributed. Thus, despite several conjectures and theorems that lend insight into the distribution of prime numbers, there is no formula to compute the next prime. The prime sieve algorithm, more formally known as the "Sieve of Eratosthenes," is a well-known and efficient technique used to find prime numbers [2].

With the growing computational needs associated with prime enumeration tasks, traditional implementations of the Sieve algorithm remain a computational bottleneck, especially given the advances in multi-core processors. As the rate of data grows and computational demands rise, harnessing the power of multithread processing becomes imperative. Over the years, there have been prior studies that have delved into the challenge of parallelizing the Sieve of Eratosthenes. Some of these approaches try to exploit the capabilities of multicore processors, breaking down the computation task into chunks [3]. This approach lends itself to using parallelization libraries and middleware, including OpenMP (shared memory) and MPI (inter processor message queues) [4]. Some of the common drawbacks and challenges of parallel implementations are that too many threads accessing shared memory can lead to memory contention, causing bottlenecks and limiting the performance gains from parallelism. Furthermore, prior work has not adequately addressed scalability issues, such that when the prime number being searched for is very large, it eventually can exceed the limits of memory. Additionally, more optimizations can be made within each iteration with improvements in mathematical algorithms, which can lead to significant savings.

We introduce a multithreaded implementation of the Segmented Sieve [3]. In our implementation, instead of handling large prime ranges in one iteration, the sieving process is broken down incrementally, which theoretically eliminates the challenges of working with large numbers, and can reduce memory usage, providing overall more efficient utilization over extended computations. In this work, we also explored some mathematical optimization to speed up computation. The code is implemented in POSIX threads C++ [6] and tested on a high-performance multi-core cluster [7]. We have been able to achieve a significant reduction in prime enumeration time, as well as a significant reduction in the associated memory footprint.

### 1.1 Sieve of Eratosthenes

Prime numbers are hard to compute, which makes them indispensable in various fields, especially in cryptography. The security of many encryption systems depends on the computational difficulty of factorizing large numbers. The "Sieve of Eratosthenes," is an old, yet efficient, algorithm used to find all prime numbers up to a specified integer. The algorithm works by iteratively marking the multiples of each prime number given a range.

The algorithm begins by creating a list of consecutive integers, from 2 up to **n**: (2, 3, 4, ..., **n**). Next, we set the current number, **P**, to the value 2. Then, remove all multiples of **P** beginning at **P**$^2$ from the list, as all multiples of **P** less than **P**$^2$ would have been removed in previous iterations already. For instance, the first time we run the algorithm, we remove all even numbers (except 2). Then, we remove all multiples of 3 (except 3), starting from 9 as 6 has already been removed, and so on. After each removal, we find the next number in the list after **P** that has not been removed. We set this new number to **P** and repeat the process.

The Sieve of Eratosthenes can be expressed in pseudocode [5], as shown in Figure 1. The Sieve of Eratosthenes is especially efficient for computing all prime numbers up to a certain limit. Its time complexity is `O(n log log n)`, making it one of the most efficient ways to find all primes up to a given limit **n**.

```
Algorithm Sieve of Eratosthenes
    input: an integer n > 1.
    output: all prime numbers from 2 through n.

    Set A[2 ... n] = true

    for i = 2, 3, 4, ..., sqrt(n):
        if A[i] is true
            for j = i², i²+i, i²+2i, i²+3i, ..., n:
set A[j] = false

    return all i such that A[i] is true.
```

**Figure 1.** Pseudocode for the Sieve of Eratosthenes.

## 2. Sieve Optimizations

Our implementation of the Sieve algorithm incorporates several optimizations to improve computational efficiency, selected specifically since we are mapping this algorithm to a multithreaded environment. We also consider a set of mathematical optimizations.

### 2.1 Memory Optimizations

Primes greater than 3 lie only in the form of 6k+1 and 6k+5, which means they are either 1 more or 5 more than a multiple of 6, as shown in Figure 2. By considering only numbers in the form 6k+1 and 6k+5, memory usage can be significantly reduced.

```
A_one:   [ 7, 13, 19, 25, 31, 37, 43, … ]
A_five:  [ 5, 11, 17, 23, 29, 35, 41, … ]
```

**Figure 2.** Prime greater than 3.

### 2.2 Divide the Range

If we are looking for primes up to a value of **n**, we can divide this range into segments. For instance, if **n**=10,000,000 and we have 10 threads, each thread would be assigned to handle 1,000,000 numbers. The main range can be split so that each thread works on contiguous segments of numbers, reducing the complexity of merging the results.

### 2.3 Multithreaded Tasks

A thread, often termed a lightweight process, represents the smallest sequence of programmed instructions that can be managed independently by a scheduler. Each thread is responsible for sieving out multiples of primes in its assigned range.

Due to memory constraints, it is not always possible to sieve very large numbers in a single thread. Utilizing consecutive chunks of memory for each thread can indeed help reduce the need for synchronization and thus can prevent memory contention. By allocating a distinct memory range for each thread, we ensure that no two threads are attempting to write to the same memory address at the same time, which is a common cause of race conditions. The segmented sieve approach sieves numbers in segments, making it memory efficient and parallel thread friendly.

### 2.4 Incremental Sieve

An incremental formulation of the sieve generates primes indefinitely (i.e., without an upper bound) by interleaving the generation of primes where the multiples of each prime **P** are generated directly by counting from the square of the prime in increments of **P**. The generation must be initiated only when the prime's square is reached, to avoid adverse effects on efficiency.

## 3 Mathematical Optimization

One of the novel contributions of our work is to create optimized mathematical operations, where each thread can work efficiently on its designated segment, ensuring that the sieve is executed in a parallelized and optimized manner.

For any thread **T**, let the current starting index in the array be **N** and the starting value be **M**, assume we want to sieve for prime number **P**.

- First, we find the greatest multiple of **P** less than **M**, denoted as **Q**.
- Then, we find **Q** *(mod 6)* = **q**, **P** *(mod 6)* = **p**, and *⌊P/6⌋* = **R**.
- Finally, we find the approximate index that a value of **Q** would be relative to **N** (this will be less than **N**).
- From there, we can go through every possible value for **p** and **q** (note that **p** can only be 1 or 5).

Let **id_1** and **id_5** be the indices to start sieving in the *mod 1* and *mod 5* arrays. Table 1 illustrates the operation performed for each **p** and **q** condition.

Table 1. Optimized Mathematical Sieve Operation

|  | q=0 | q=1 | q=2 | q=3 | q=4 | q=5 |
|---|---|---|---|---|---|---|
| **p=1** | id_1 += R<br><br>id_5 += 5*R | id_1 += (6*R+1)<br><br>id_5 += 4*R | id_1 += (5*R+1)<br><br>id_5 += 3*R | id_1 += (4*R+1)<br><br>id_5 += 2*R | id_1 += (3*R+1)<br><br>id_5 += 1*R | id_1 += (2*R+1)<br><br>id_5 += (6*R+1) |
| **p=5** | id_1 += 5*(R+1)-1<br><br>id_5 += R | id_1 += (6*(R+1)-1)<br><br>id_5 += (2*(R+1)-1) | id_1 += (R+1)<br><br>id_5 += (3*(R+1)-1) | id_1 += 2*(R+1)<br><br>id_5 += (4*(R+1)-1) | id_1 += 3*(R+1)<br><br>id_5 += (5*(R+1)-1) | id_1 += 4*(R+1)<br><br>id_5 += (6*(R+1)-1) |

Once we know the starting index, we can iterate by adding **P** to the index (essentially adding 6*P to the number). This guarantees that the new index will be the same remainder mod 6 and that it will be a multiple of **P**.

**Example:** Assume that we are using a chunk size of 6,000 values and using 5 threads (0-4). Also assume that we are currently iterating through the group 12,000-18,000, which is mapped onto thread 2 of 5. The index range would be 400-599, and we would sieve from 14,400 – 15,599. Let us sieve for the prime 13. The value of **R** would be 2. The smallest multiple of 13 less than 14,400 is 14,391. *14,391 mod 6 = q = 3*, and *13 mod 6 = p = 1*. id_1 and *id_5* would be 400-2 = 398, as 14,391 would be between those two values (14,389 and 14,393). *id_1* = 398 + 4*2 + 1 = 407, and *id_5* = 398 + 2*2 = 402. Converting these numbers back to integers, we see that we start sieving from 407*6 + 1 + 12,000 = 14,443 and 402*6 + 5 + 12,000 = 14,417, which we can confirm are multiples of 13 and are 1 and 5 mod 6, respectively.

This method reduces redundant calculations, simplifies each step's mathematical operation, and ensures that each thread is not just parallel but also efficiently parallel, making the best use of the computational resources.

## 4.0 Implementation

In Figure 3, we provide a diagram of our multithreaded Sieve algorithm's execution flow when utilizing multiple threads. We show the thread structure and flow of the information. In Figure 4 we provide pseudocode for our implementation. In each iteration, the newly found prime numbers are stored in the Prime Number Data Store, which we currently store in a linked list. The primes are also stored in the Sieve Data Store, but only up to the square root of the numbe,r since we only need to sieve up to the square root of the number. This is because for any prime greater than the square root, $P^2$ would be greater than $N$ and there would be nothing to sieve.

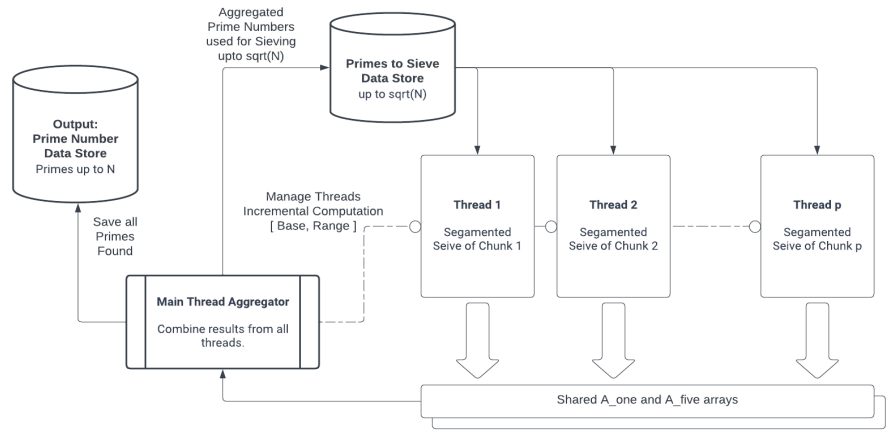

**Figure 3.** Parallel execution of our multithreaded Sieve algorithm.

```
Algorithm: Multi-thread Sieve

Thread: Main Thread
    Input: Threads t, Number to sieve to N, Segment length S

    Set mod1 [1, 7, 13, 19 ... a<N] = true
    Set mod5 [5, 11, 17, 23 … b<N] = true
    for i = 0, 1, 2 … N/S-1:
        for j = 0 … t-1
            Create Worker Thread j
                Worker Thread j sieve (start=i*S, end=i*S + j*S/t)
        return all k such that mod1[k] = true or mod5[k] = true

Thread: Worker Thread
    Input: start, end

    Set A = current list of primes
    for i in A:
        Compute starting indexes
        Sieve until end using Optimized Mathematical Sieve Operation
```

**Figure 4.** Pseudocode for our implementation.

## 5. Results and Discussions

Our Sieve algorithm implementation uses the POSIX threads C++ libraries [7]. The performance study was carried out on the system described in Table 2. By leveraging a high-performance computing cluster, our implementation aims not just to sieve primes efficiently, but to push the boundaries of what's possible in large-scale prime enumeration.

**Table 2. System details**

| Number of cores | Memory size (MB) | C++ version | Operating System |
|---|---|---|---|
| 24 (used 3) | 128000 | 11 | CentOS Linux 7 (Core) |

The computation time is measured while varying the number of threads and the value **N**, which specifies the range that we are sieving to. The memory usage is 1,000,000 x 4B x 2 = 8MB for the two **A_one** and **A_five** arrays.

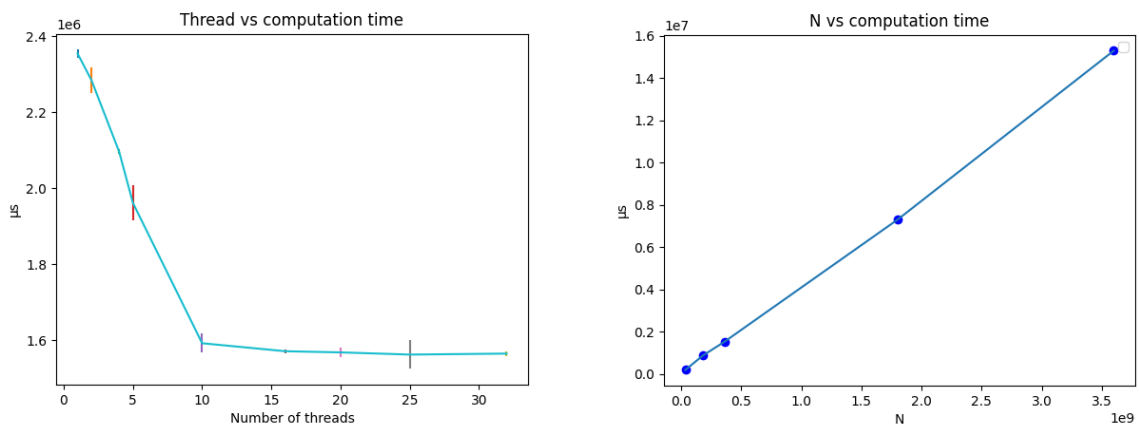

**Figure 5.** Computation time while varying the number of threads (left) and varying the value of **N** (right).

Inspect the graph on the left, the computation time drops linearly against the number of threads until thread number = 10. In the right graph, we can see the linear correlation between the value of **N** and computation time. The result shows that our algorithm is highly scalable.

The interplay between memory size and computational speed is a pivotal aspect of algorithm optimization. Our second set of experiments delved deeper by exploring how the sizes of the **A_one** and **A_five** arrays, which serve as the scratch memory during each thread's operations, influence the algorithm's overall performance. When the sizes of the scratch memory arrays increased, the number of iterations the program needs to complete its task decreased. Using a larger memory pool, more prime sieving results can be held at once, enabling more extensive sieving in a single pass. Thus, fewer passes or iterations are required to achieve the end goal. Our experiments ranged from using a tiny scratch memory size of 100k, suitable for environments with stringent memory restrictions, to a spacious 100M, tailored for systems where memory is abundant. The goal was to find a balance - a size that minimizes the iteration count, while also being mindful of memory usage constraints.

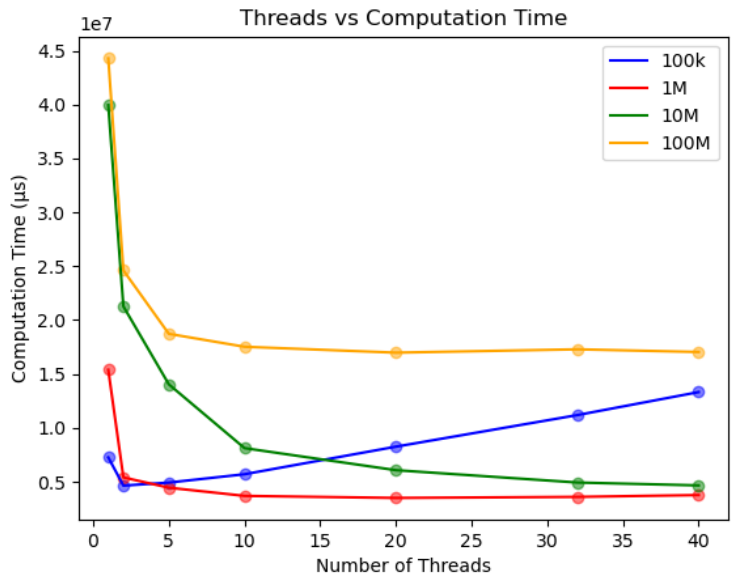

**Figure 6.** Computation Time for Various Array Sizes

The results from these experiments have shown some very interesting findings. At first glance, one might assume that a scratch memory size of 100M, the largest in our experiments, would achieve the best performance in terms of computation time. However, the performance was poor when using this configuration. We believe it is an attribute of the intrinsic memory constraints of what the parallel node can allocate efficiently. When the memory exceeds this limit, the operating system swaps pages, which is causing significant delays. This swapping of memory pages between the RAM and the disk is notoriously slow, introducing significant delays in execution. Hence, while having a large scratch memory may reduce iterations, the overheads introduced due to page swapping can negate those benefits.

For the 100k array size, where each thread uses the least amount of memory, we ran into different challenges. The execution time grew as the number of threads increased, an anomaly unique to this configuration. A higher thread count implies that each thread's designated segment shrinks. When these segments approach, or dip below, the size of a cache line, cache synchronization becomes mandatory among threads. This synchronization can lead to stalls, as cache lines are evicted and fetched repeatedly, impacting performance.

The best performance is achieved by 1M array size, as this amount of scratch memory can avoid cache collisions and page swapping. Overall, the cache and memory play a significant role in the overall performance. Choosing the right scratch memory size is a critical parameter to achieve good multithread performance.

# 6 Conclusions

In this paper, we explored how to better parallelize the execution of the Sieve of Eratosthenes. Clearly, both the arithmetic complexity of computation and the storage requirement play a big role in the computation speed of the standard implementation. However, our proposed algorithm is faster and more compact (requiring **N**/3 auxiliary storage), resulting in a more efficient parallel execution of the Sieve of Eratosthenes. We explored selecting the chunk size to optimize memory usage, which can avoid memory thrashing. Our incremental segmented approach enables us to handle huge numbers, almost approaching infinity. This is crucial for large-scale prime number computations and applications that require extensive use of primes.

Memory management in high-performance multi-threaded computations is extremely important. The cache sizes and physical memory constraints can have a huge impact on performance. We encountered this when selecting the size of our scratch memory. What we found is that there are tradeoffs in terms of selecting this size.

Next Steps for advancing the Multithreaded Sieve Algorithm Implementation:

1. Prime Data Store File Writing: The current setup is poised for indefinite expansion. However, to efficiently manage and store results, there is a need to implement writing to files, ensuring the preservation of computed data. To address potential memory constraints, the prime base list will be segmented. Each segment will contain up to 125 million primes to prevent memory overflow and ensure efficient retrieval.
2. Custom Big Integer Library: In order to sieve infinitely, a special library for large numbers needs to be created. While there are default big integer libraries, we can customize our own for further optimization. For example, we only divide by 6, so a special operation can be created. Furthermore, a special operation can be created to efficiently mod 6.

3. Incorporation of Multiprocessing: To further optimize the computational process, the implementation of multiprocessing will be explored. This approach is to divide the tasks into multiple computation nodes. However, this comes with its set of challenges. One primary concern is minimizing data transfer between processes, ensuring coherence and consistency in results.

## Source Code

GitHub: https://github.com/evanning1/Multi-thread_Prime_Sieve

## References

[1] R.L. Rivest, A. Shamir, L. Adleman, "A Method for Obtaining Digital Signatures and Public-Key Cryptosystems," (1978) Massachusetts Institute of Technology, https://people.csail.mit.edu/vinodv/COURSES/MAT302-S12/Rsapaper.pdf

[2] Xuedong Luo, "A Practical Sieve Algorithm Finding Prime Numbers," (1989) Communications of the ACM, 32(3), March 1989, 344–346. https://doi.org/10.1145/62065.62072

[3] Torben Hansen, "Parallel Implementation of the Sieve of Eratosthenes," F120116 Utrecht University - Institute of Mathematics, https://himsen.github.io/pdf/Project1_parallel_algorithms_Torben.pdf

[4] Mário Cordeiro, "Parallelization of the Sieve of Eratosthenes, Faculdade de Engenharia da Universidade, s/n 4200-465 Porto PORTUGAL .

[5] Jonathan Sorenson, "An Introduction to Prime Number Sieves," Computer Sciences Technical Report #909, Department of Computer Sciences University of Wisconsin-Madison, January 2, 1990.

[6] POSIX Threads Programming https://hpc-tutorials.llnl.gov/posix/

[7] Research Computing at Northeastern University https://rc-docs.northeastern.edu/en/latest/